\documentclass[twocolumn]{webofc}

\usepackage[varg]{txfonts}   
\usepackage{hyperref}
\usepackage{url}
\hypersetup{colorlinks=true,citecolor=blue,urlcolor=blue,linkcolor=blue}
\usepackage{graphicx}    
\usepackage{subcaption} 
\usepackage{amsmath,amssymb}

\newcommand{\myparagraph}[1]{ {\noindent{\textbf{#1}}}}

\newcommand{\vishali}[1]{\color{black}{#1}}

\begin{document}

\title{Topological signatures of jamming in granular force networks via persistent homology}

\author{
    Vishali S$^{1}$\fnsep\thanks{\email{vishalis@iisc.ac.in}}
    \and
    Abrar Naseer$^{2}$
    \and
    Vijay Natarajan$^{3}$
    \and
    Tejas G Murthy$^{2}$
}

\institute{
  Department of Mathematics, Indian Institute of Science, Bengaluru, 560012, Karnataka, India
  \and
  Department of Civil Engineering, Indian Institute of Science, Bengaluru, 560012, Karnataka, India
  \and
  Department of Computer Science and Automation, Indian Institute of Science, Bengaluru, 560012, Karnataka, India
}

\abstract{
Granular materials exhibit intricate force networks, often concentrated in chains rather than being uniformly distributed. These meso-scale structures are complex to characterize, owing to their heterogeneous and anisotropic nature. We present a \vishali{topological analysis} in which we characterize force networks by studying the connectivity of contact forces in granular materials. We obtain the force network data during the isotropic compression of a 2D granular ensemble, which comprises of bidisperse photoelastic disks. The force chains are visualized using a bright-field polariscope, which are then analyzed using photoelasticity techniques to get contact force information across all the particles in the granular ensemble. The forces and contact network are studied using topological descriptors. We analyze the evolution of these descriptors using methods developed in the areas of computational topology and persistent homology. Specifically, we evaluate the number of homology class generators to capture critical changes in network connectivity with the evolution of packing fraction. \vishali{Building on previous applications of persistent homology, our topological analysis of the force network provides valuable insights by distinguishing different phases of particle interactions and revealing unique structural transitions around jamming and thus help enhance our understanding of the stability and jamming dynamics in granular systems.}
}

\maketitle


\section{Introduction}
\label{sec:intro}
\vspace{-0.5em}
Granular materials exhibit complex spatial organization in the way they transmit forces, often through highly localized structures known as ``force chains'' These structures play a critical role in determining mechanical response, stability and transitions such as jamming. However, capturing the evolving organization of this network of force chains - i.e. the force network, particularly at and around the jamming transition, remains a fundamental challenge.

One of the earliest experimental approaches to directly visualize vector contact forces was introduced by Majmudar and Behringer~\cite{majmudar2005contact}, who used photoelastic particles to infer vector contact forces in 2D granular packings. This technique was further refined and reviewed comprehensively by Daniels \textit{et al.}~\cite{daniels2017photoelastic}, who detailed the methodology of force reconstruction using birefringent materials and polarized light imaging. These advances laid the foundation for experimental quantification of force networks.

On the computational side, topological techniques have emerged as powerful tools to analyze force networks. Arévalo \textit{et al.}~\cite{arevalo2010topology} used polygonal analysis of force-chain structures to study transitions in network connectivity near jamming. Building on this approach, Kondic \textit{et al.}~\cite{kondic2012topology} introduced the use of Betti numbers to quantify the number of connected components ($\beta_0$) and loops ($\beta_1$) in force networks, demonstrating how topological invariants evolve under compression. \vishali{Kramár \textit{et al.}~\cite{kramar2014quantifying} described an extension to persistent homology and applied it on three different simplicial complexes defined on the granular force network.} A significant milestone in this domain was the comprehensive review by Papadopoulos \textit{et al.}~\cite{papadopoulos2018network}, which surveyed network-based approaches to studying particulate systems. This work synthesized developments across experimental, simulation, and theoretical fronts, highlighting the growing relevance of topological data analysis (TDA) in understanding multiscale features in granular media up to 2017.

\vishali{Basak \textit{et al.}~\cite{basak2021two} compared force networks constructed from different photoelastic reconstruction methods, concluding that persistent homology provides robust comparative metrics across different reconstruction approaches.} In subsequent work, Basak \textit{et al.}~\cite{basak2023evolution} analyzed the temporal evolution of force networks during stick-slip motion of an intruder in a granular medium, revealing how the birth and death of topological features can signal impending slip events.
\begin{figure*}[!ht]
  \centering

  \begin{subfigure}{0.19\textwidth}
    \includegraphics[width=\linewidth]{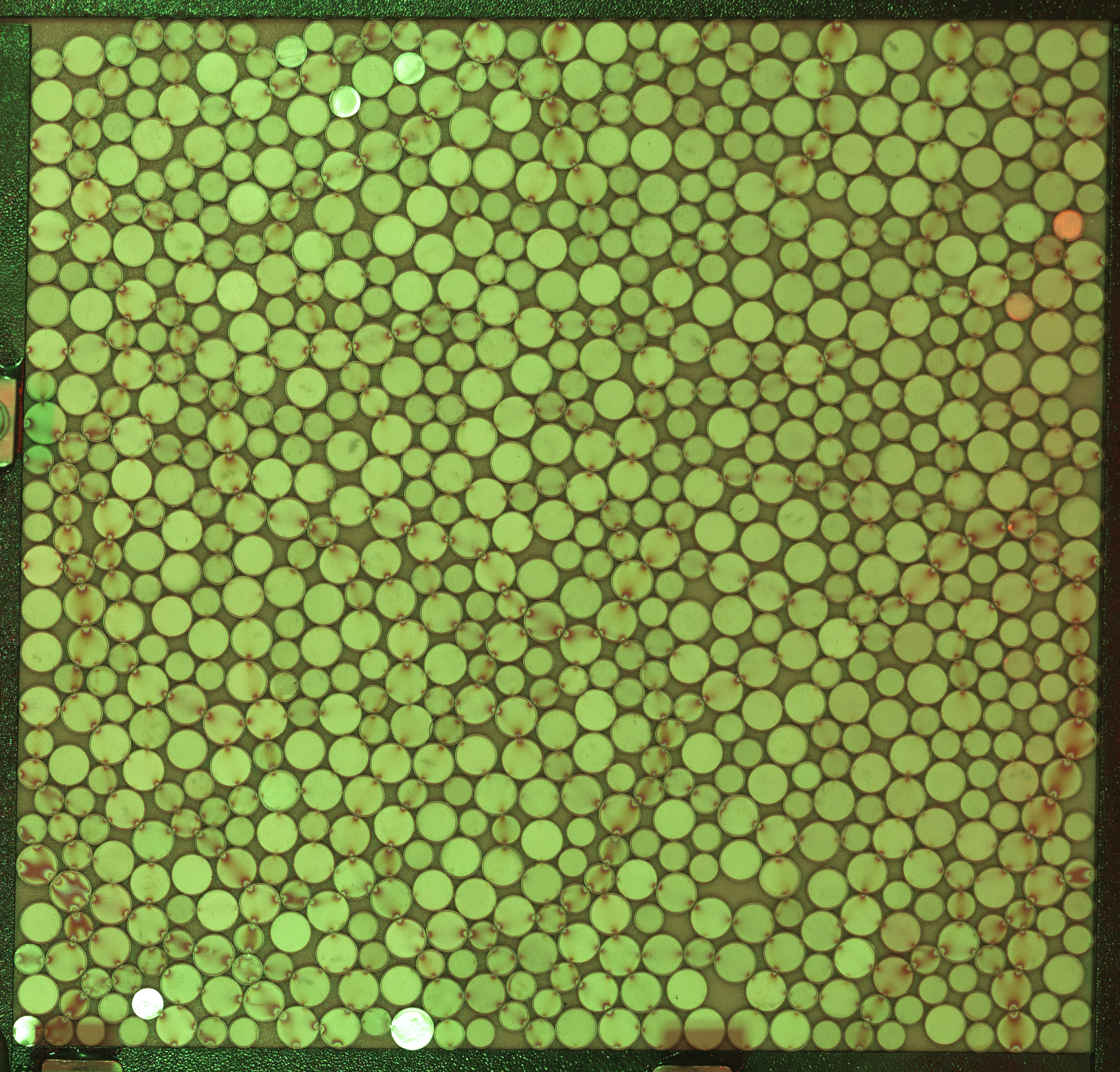}
    \caption{}
    \label{fig:panel-a}
  \end{subfigure}
  \hfill
  \begin{subfigure}{0.15\textwidth}
    \includegraphics[width=\linewidth]{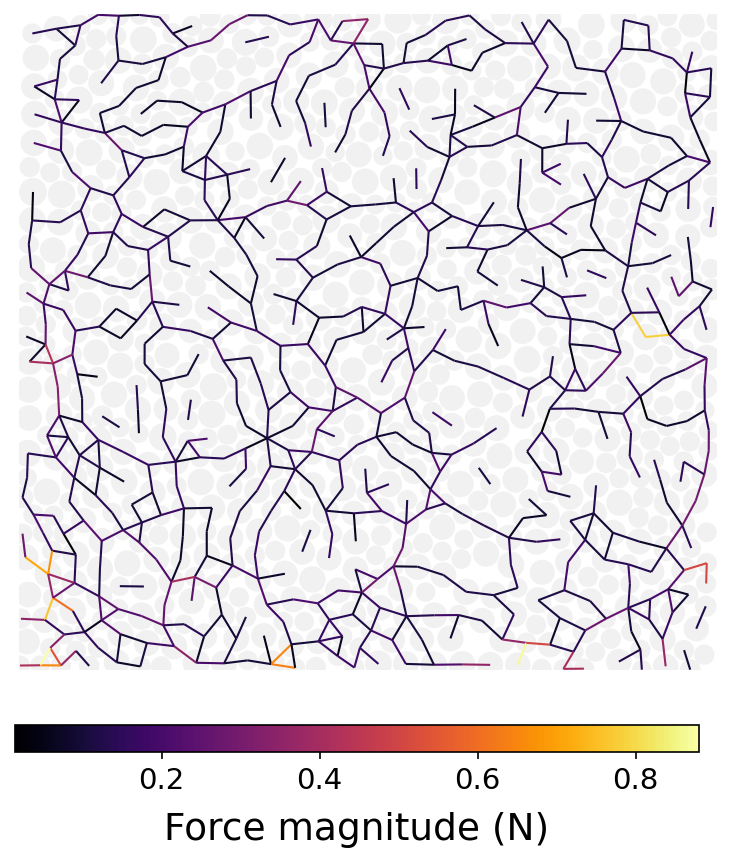}
    \caption{}
    \label{fig:panel-b}
  \end{subfigure}
  \hfill
  \begin{subfigure}{0.19\textwidth}
    \includegraphics[width=\linewidth]{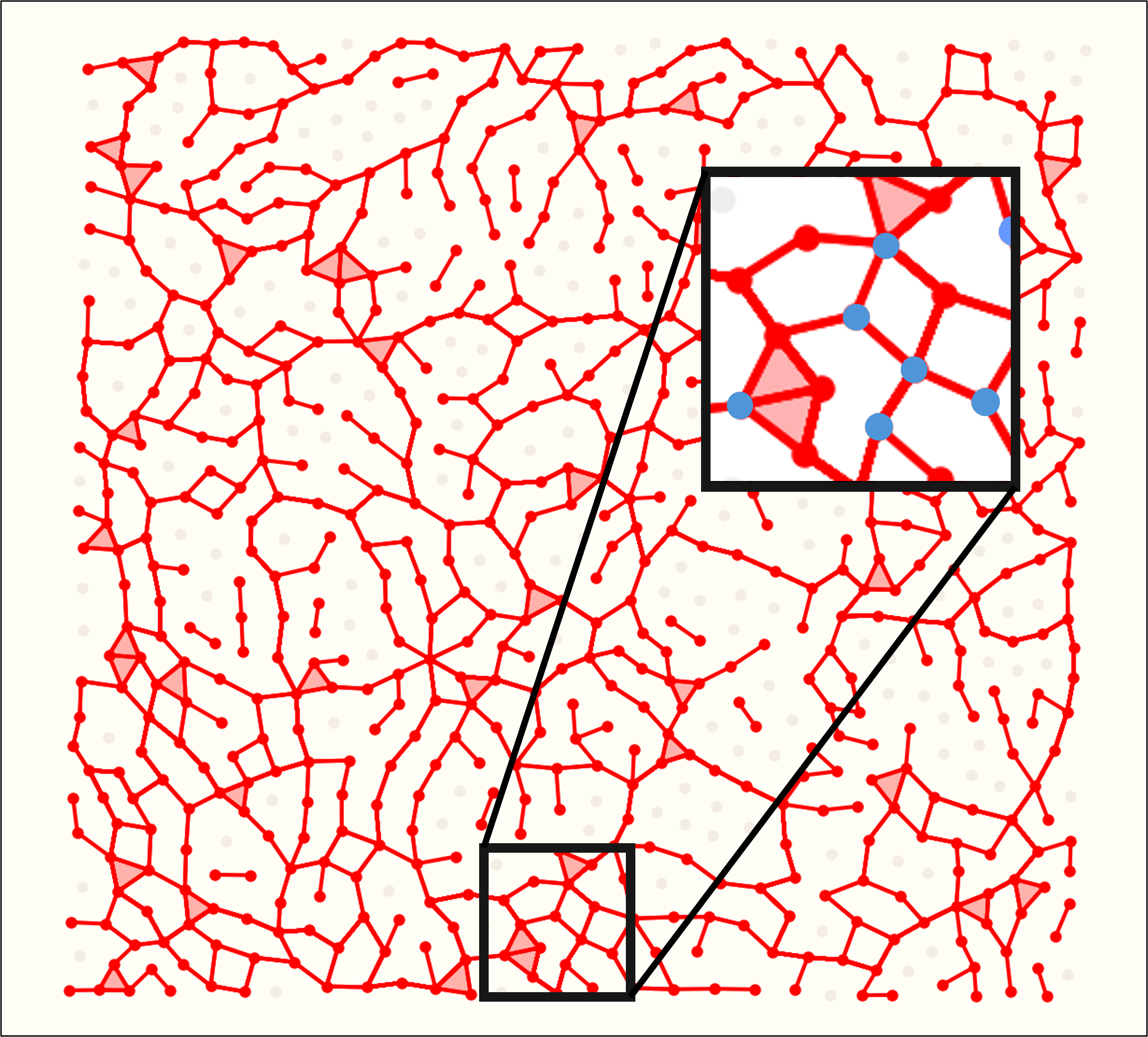}
    \caption{}
    \label{fig:panel-c}
  \end{subfigure}
  \hfill
  \begin{subfigure}{0.20\textwidth}
    \includegraphics[width=\linewidth]{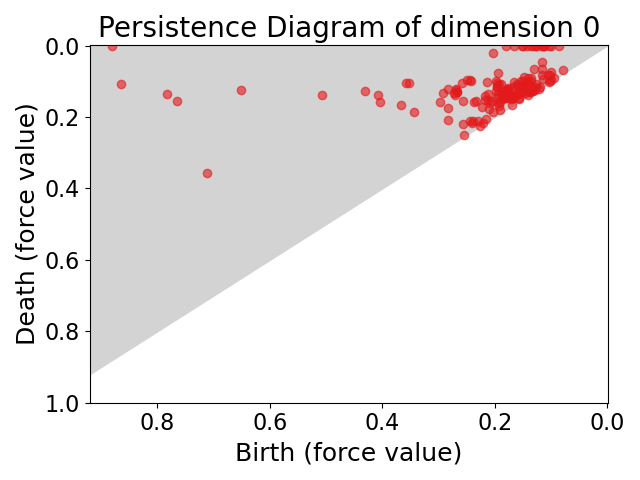}
    \caption{}
    \label{fig:panel-d}
  \end{subfigure}
  \hfill
  \begin{subfigure}{0.20\textwidth}
    \includegraphics[width=\linewidth]{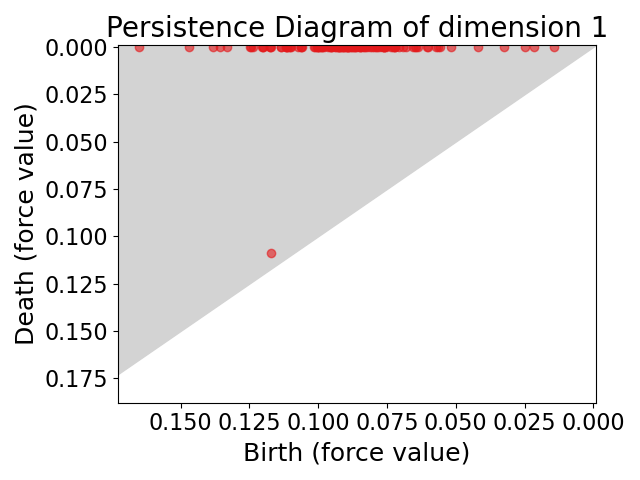}
    \caption{}
    \label{fig:panel-e}
  \end{subfigure}
  \caption{A pipeline for network construction and analysis of the photoelastic experimental data. (a)~Input photoelastic image in green channel at $\phi = 0.77897$. (b)~Interparticle force lines \vishali{having force magnitudes below 10 times the average force,} extracted from the experimental data. (c)\vishali{~Simplicial complex constructed from all the contact forces under the aforementioned threshold (red dots, lines and triangles are used to show the 0, 1 and 2 dimensional simplices). A close-up view of the locations of some 0-th homology class generators (light blue dots) inside a particular square region.} (d,e)~Persistence diagrams of 0- and 1-dimension, respectively.}
  \label{fig:panel}
  \vspace{-0.5em}
\end{figure*}

\vishali{The present study advances prior work by applying persistent homology to force networks reconstructed from photoelastic data during isotropic compression. Rather than solely classifying persistent homology features to characterize the particle systems, we examine how these features evolve in relation to internal changes within the network. This enables us to trace the rise and spread of localized high-force regions, leading to clear early signatures of the jamming transition.}

\section{Experimental Setup}
\label{sec:exp}
\vspace{-0.5em}
We construct a birefringent 2D granular ensemble by cutting disk-shaped particles from a photoelastic sheet (Vishay PSM-4). The ensemble consists of a bidisperse system made up of 832 particles (See figure~\ref{fig:panel-a}), with diameters $D_1$ = 15.4 mm and $D_2$ = 11 mm, in a $\approx$1:1 concentration ratio \vishali{in terms of particle number}. The ensemble is confined by rails within an area $\approx$ $0.4$~m $\times 0.4$~m, where two movable rails are operated by a linear actuator (12v, 2 Amps) to apply biaxial compression at a speed of 0.38~mm/s. The particles rest on a porous sheet that provides air cushioning, minimizing base friction between the particles and the sheet, which ensures that interactions are solely particle-particle. To implement photoelasticity, we construct a brightfield polariscope to leverage the birefringent property of the particles to extract vector contact force information. The camera used for imaging has a resolution of 134.5 microns and captures images at 1 frame per second (fps). This vector contact force data is used to analyze the topological features of the granular ensemble.
We make use of PeGS (Photoelastic Grain Solver - an open source software) \cite{Kollmer2024jekollmer}, which follows a systematic approach to convert the experimental images into force networks as summarized below:

\noindent \textbf{Force calculation:} 
We process each experimental image to detect particles and force-bearing contacts. 
Each force-bearing particle is analyzed to evaluate the vector contact forces by solving the inverse problem \cite{majmudar2005contact,daniels2017photoelastic}. The outcome is a weighted adjacency matrix that encodes the contact forces between all pairs of particles in contact at a given compression step.

\noindent \textbf{Thresholding:} 
\vishali{A force threshold ($f^*$) applied to this matrix filters out contacts with force magnitudes above 10 times the mean force, to eliminate noisy contacts. All forces below this threshold, however small, are considered for all analysis in the following sections} (See figure~\ref{fig:panel-b}).

\section{Methodology}
\label{sec:topo}
\vspace{-0.5em}
\myparagraph{Topological descriptors.}
Topological data analysis (TDA) provides a set of tools for extracting the global structure from complex datasets. A key concept is \textit{homology}, which characterizes a network represented as a \textit{simplicial complex}~\cite{edelsbrunner2010computational} by measuring its \textit{holes} (features) in different dimensions. \textit{Persistent homology} extends this by considering the lifetime of an individual feature as one varies a filtration parameter. This analysis results in the construction of a \textit{persistence diagram} (Figures~\ref{fig:panel-d} and ~\ref{fig:panel-e}) that enables the tracking of \textit{birth} and \textit{death} of significant topological features within the network while varying the filtration parameter. Sharp transitions in the diagram may signal a critical rearrangement or a new structural phase in the granular packing. A point in the persistence diagram is called a \textit{persistence pair} and the associated birth is located at a \textit{generator}.

\myparagraph{Computation.}
A simplicial complex is constructed from the force network by adopting the \emph{interaction network} model~\cite{kramar2014quantifying}, where each particle is represented as a vertex and the inter-particle contact force is represented by an edge. Higher dimensional simplices (triangles, tetrahedra) are included when all lower-dimensional faces belong to the network (Figure~\ref{fig:panel-c}). The filtration parameter at an edge is assigned equal to the contact force, at a vertex is set to the maximum over all incident edges, and at a triangle to be equal to the minimum over all bounding edges~\cite{kramar2014quantifying}. We compute the following using the \texttt{GUDHI} library~\cite{gudhi2015gudhi}:
\begin{itemize}
\item \textbf{$\beta_0$} - zeroth Betti number, captures the number of connected components, indicating groups of particles in force-carrying clusters.
\item \textbf{$\beta_1$} - first Betti number, captures the number of loops or holes, highlighting the presence of circuits in the force network.
\item \textbf{$\gamma_0$} - the number of homology generators in the persistence diagram of dimension-0 (connected components).
\item \textbf{$\gamma_1$} - the number of homology generators in the persistence diagram of  dimension-1 (loops).
\end{itemize}
When tracked over the compression sequence, the above quantities reveal the evolution of the network near the jamming transition.\vishali{The analysis focuses on $\beta_0$, $\beta_1$, $\gamma_0$ and $\gamma_1$ since the input is represented as a 2D simplicial complex.}

\section{Results and Discussions}
\label{sec:results}
\begin{figure}[htb]
\centering
\includegraphics[width=7cm,clip]{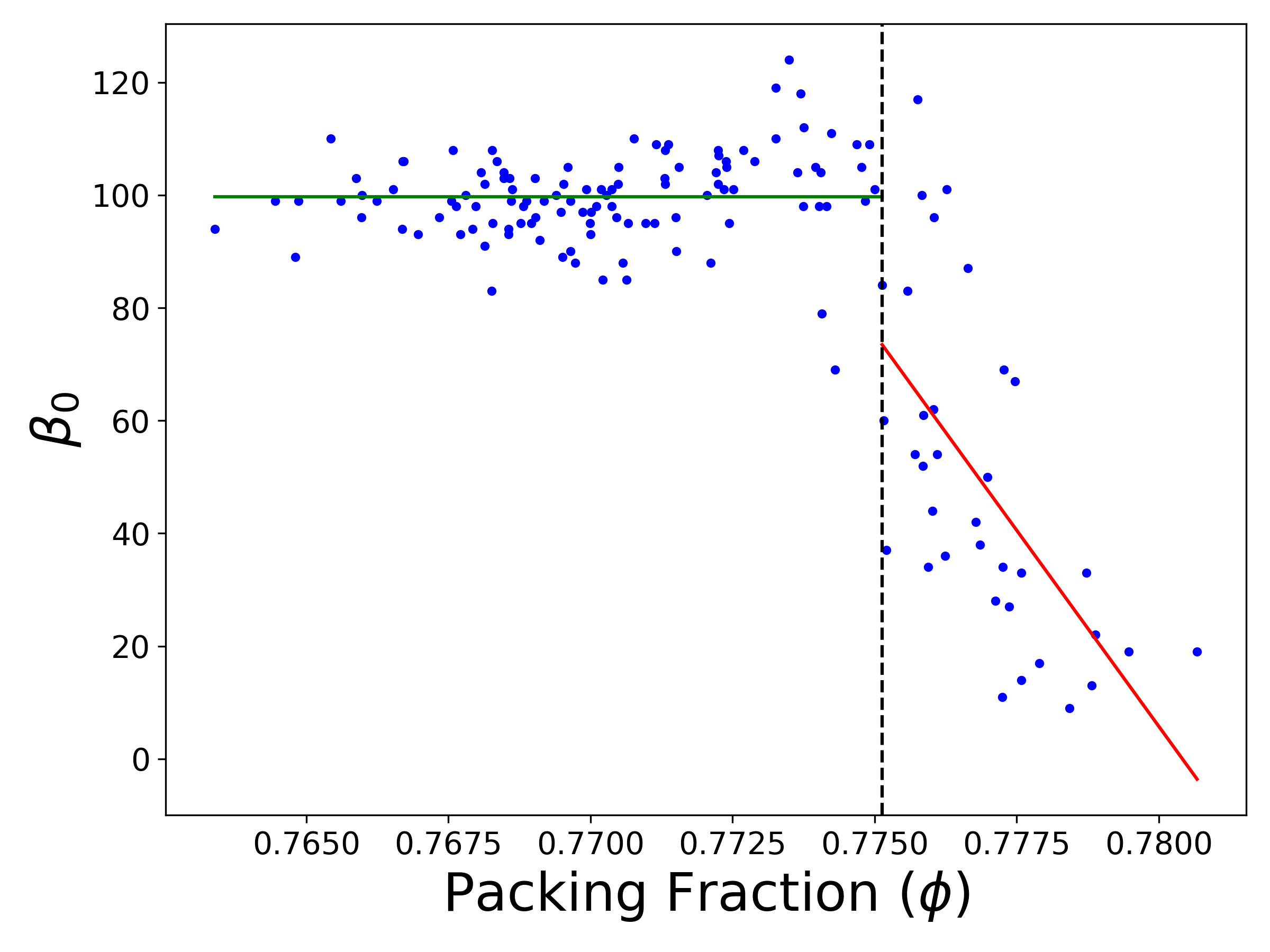}
\includegraphics[width=7cm,clip]{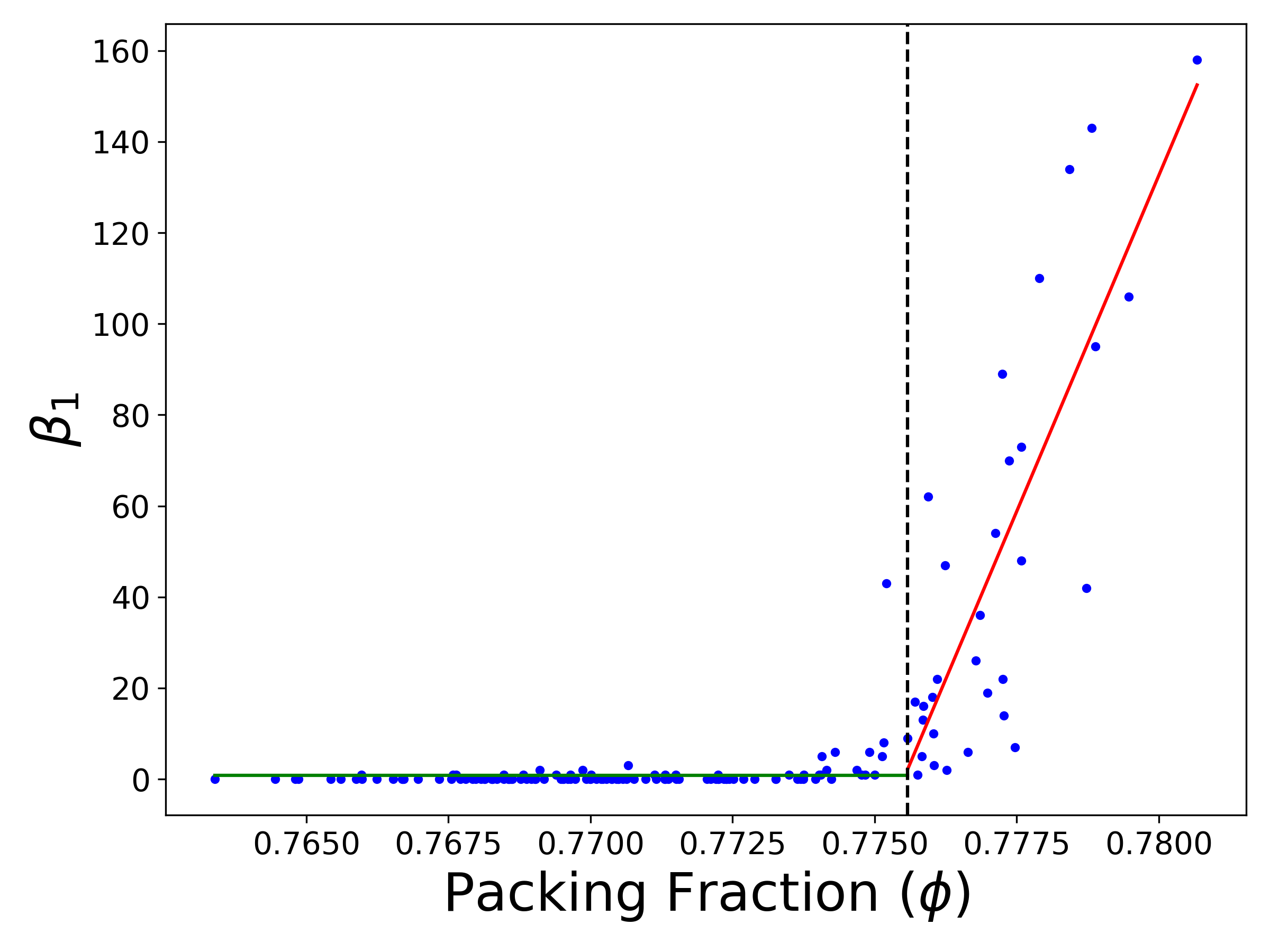}
\caption{Variation of $\beta_0$ and $\beta_1$ with packing fraction $\phi$.}
\label{fig:betti-number-variation}
\vspace{-0.5em}
\end{figure}

\vspace{-0.5em}
We observe that during the early stages of compression, the force network exhibits numerous short-lived loops that emerge and quickly disappear. As the system approaches jamming, certain loops and clusters become more robust, indicating the formation of stable pathways for force transmission. The persistence diagrams highlight these observations, with long-lived loops signifying crucial structural features within the force network. Additionally, as we approach jamming, particles group into force-carrying clusters, highlighting the buildup of a single network-spanning force backbone, often regarded as the signature of a jammed state. Figure~\ref{fig:betti-number-variation} shows that $\beta_0$ undergoes a sharp drop while $\beta_1$ exhibits a pronounced spike around packing fraction $\phi \approx 0.7755$. The transition points of change in the trend of the descriptors are identified through clustering by fitting two piecewise linear segments, with the first segment constrained to be horizontal, to highlight the onset of structural change. This behavior is indicative of a structural transition occurring due to the emergence of loop-rich connectivity patterns—signatures of mechanically stable, load-bearing configurations. \vishali{The transition seen in $\beta_0$ and $\beta_1$ seem to be in agreement with the physical definition as well as the previous studies of jamming~\cite{kondic2012topology, kramar2014quantifying}.}

\begin{figure}[htb]
\centering
\includegraphics[width=7cm,clip]{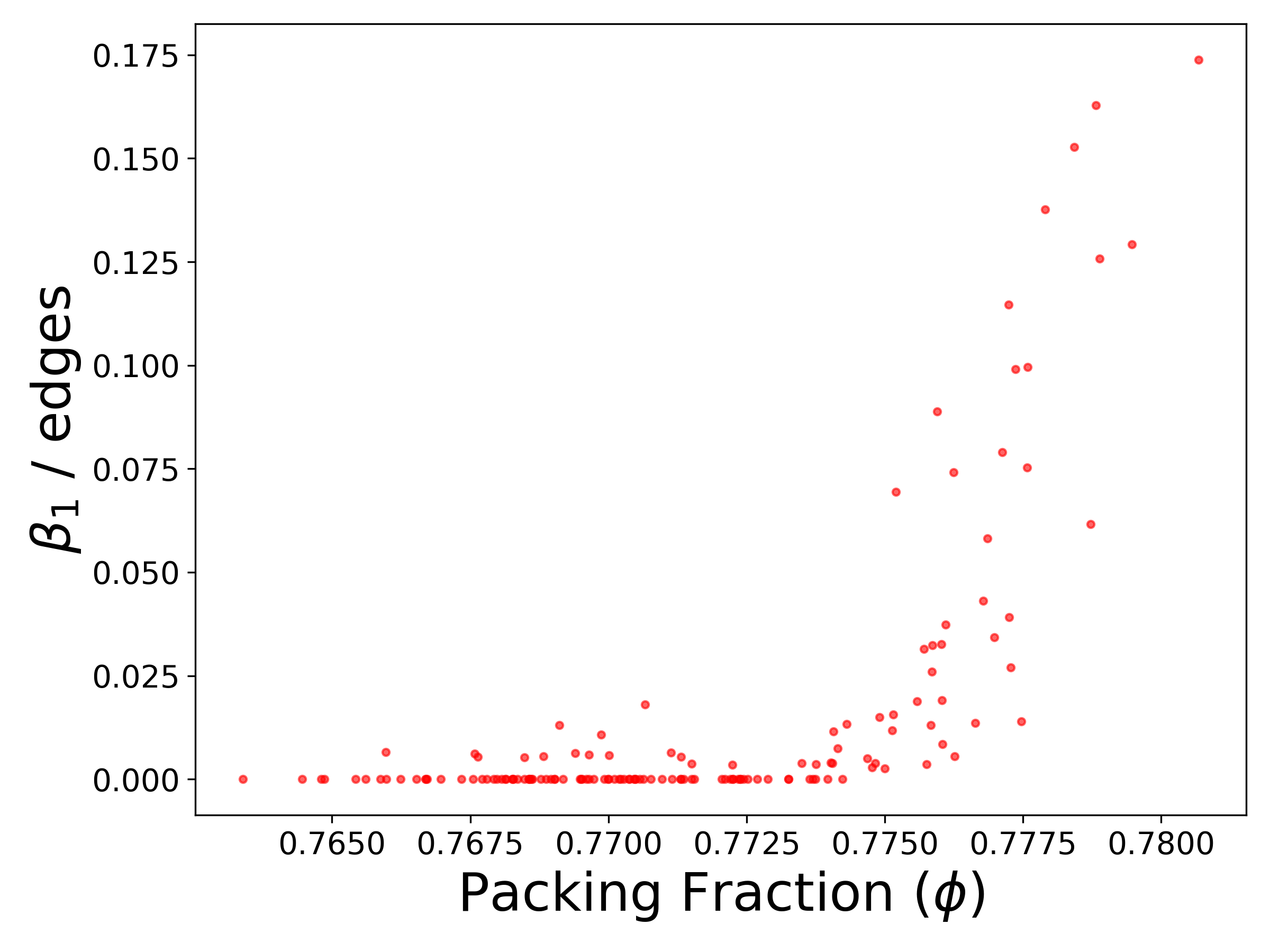}
\caption{$\beta_1$ normalized by the total number of edges in the force network, plotted against packing fraction $\phi$.}
\label{fig:normalized-beta1-variation} 
\vspace{-0.5em}
\end{figure}

Figure~\ref{fig:normalized-beta1-variation} shows the evolution of $\beta_1$/$\#edges$ with increasing packing fraction. We observe that beyond the transition point ($\phi \approx 0.7755$), the number of loops grows more rapidly than the total number of force-bearing contacts (edges). Rather than forming isolated branches, contacts begin to reinforce local rigidity by building nested or overlapping cycles, a hallmark of robust jammed packing, thereby increasing the count of independent cycles in the network. From a topological standpoint, this behavior indicates a denser and more interconnected force network, where loop proliferation is a dominant mechanism of structural evolution. Notice that this transition starts well before $\phi \approx 0.7755$ and around $\phi \approx 0.774$. 

\begin{figure}[htb]
\centering
\includegraphics[width=7cm,clip]{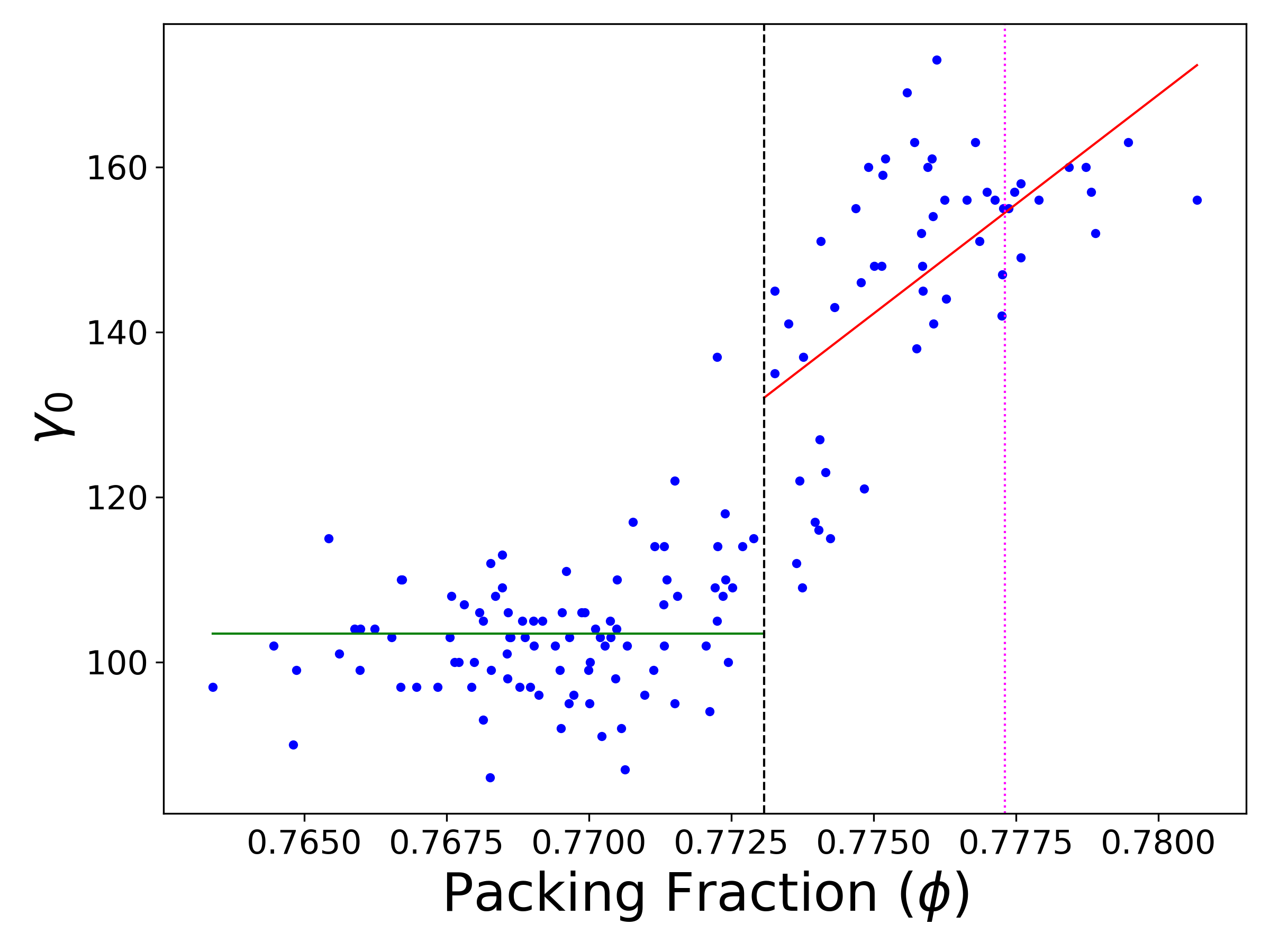}
\caption{Variation of $\gamma_0$ with packing fraction $\phi$. \vishali{After $\phi = 0.7773$ (dotted magenta line), the $\gamma_0$ value stays saturated within a band of 160$\pm{10}$.}}
\label{fig:gamma0-variation} 
\vspace{-0.5em}
\end{figure}

While $\beta_0$ and $\beta_1$ provide useful markers of this transition, they are not uniquely descriptive of the force network itself. These values are invariant to both the filtration method employed and the size of the network. For example, a fully connected network that contains a single loop of 10 (smaller loop) or 100 particles (larger loop) would yield the same $\beta_1$ value, despite having vastly different internal
structures.

Thus, the Betti numbers alone—though informative of the transition—do not capture the complete structural richness of the force network. Additional descriptors, such as the number and persistence of homology generators, are required to quantify the topological complexity in a size- and scale-sensitive manner. We observed that $\gamma_1$ and $\beta_1$ are similar and differ only by the number of triangles in the simplicial complex, and hence focus on $\gamma_0$. Figure~\ref{fig:gamma0-variation} shows the evolution of $\gamma_0$ as a function of packing fraction.  $\gamma_0$ provides a more sensitive measure of the internal structure of the force network than  $\beta_0$, particularly in the context of jamming. Initially, as the system approaches the jamming threshold, the rise in $\gamma_0$ closely tracks that of $\beta_0$, suggesting that each generator corresponds to a distinct connected component of force-bearing particles. However, beyond the packing fraction $\phi \approx 0.7732$, we observe a rapid increase in $\gamma_0$ despite the continuing decrease in $\beta_0$. This divergence indicates that although the number of macroscopic components decreases (as clusters merge), the internal structure of those clusters becomes increasingly intricate. In particular, between $\phi \approx 0.7732$ and $\phi \approx 0.7773$, $\gamma_0$ rises significantly before saturating, revealing that the network begins to exhibit a high density of short-lived, low-persistence generators (birth - death force value less than 0.1 N), contributing to increase and saturation in the \textit{roughness}~\cite{kramar2014quantifying} of the network. 

At low packing fractions, these force-bearing particles are sparsely distributed, and the force chains they form tend to be isolated, resulting in a sparser network with fewer generators. As the system transitions into the jammed regime, these previously disconnected force chains begin to merge. This leads to the formation of new topological features (births), followed by rapid merging (deaths), particularly of components with lower persistence. The saturation in $\gamma_0$ post $\phi \approx 0.7773$ suggests that the force network reaches a structurally stable regime, where most topologically significant components have already formed and merged. This transition observed in $\gamma_0$ is consistent with the critical packing fraction $\phi_{c} \approx 0.7730$ reported in~\cite{Abrar2025} on this very same experimental data, where the onset of jamming was identified based on average force per particle.

 This close agreement further supports the interpretation that the increase and saturation of $\gamma_0$ reflect the development of a mechanically stable, system-spanning force network. Its sensitivity to both the number and lifespan of features makes it a more informative descriptor for identifying the onset of jamming and analyzing the evolving structural complexity of the force network.

\section{Conclusions}
\label{sec:conc}
\vspace{-0.5em}
In this work, we have demonstrated the utility of topological descriptors in capturing the structural evolution of force networks in a granular ensemble undergoing isotropic compression.

\vishali{Betti numbers provided useful global markers: a sharp drop in $\beta_0$ and a concurrent spike in $\beta_1$ near $\phi \approx 0.7755$ signaling a topological transition around the jamming phase. However, these descriptors could not fully capture the complexity of the network because they are invariant to system size and the filtration.}

In contrast, $\gamma_0$ -- particularly those with low persistence -- offered a higher sensitivity to internal structural transitions. We observed that $\gamma_0$ rises sharply beginning at $\phi \approx 0.7732$ and saturates near $\phi \approx 0.7773$, reflecting a phase where isolated force network merge and generate transient topological features. \vishali{This behavior suggests that $\gamma_0$ is not only sensitive to network reorganization but also predictive of mechanical stability and acts as an early marker to loop proliferation and jamming transition}. Its evolution captures both the formation and dissolution of force-carrying clusters—making it a finer and more dynamic measure of structural complexity. \vishali{The methods used here naturally extend to three-dimensional granular systems by incorporating 3D simplices into the interaction network. This approach can also aid in analyzing pore structures within 3D packings, offering insights into connectivity and void space geometry.}
\vspace{-1em}
\bibliography{final}
\end{document}